\documentclass[conference]{IEEEtran}
\IEEEoverridecommandlockouts
\usepackage{multirow}
\usepackage{subfigure}
\usepackage{float}
\usepackage[pdftex]{graphicx}
\usepackage{cite}
\usepackage{amsmath,amssymb,amsfonts}
\usepackage{algorithmic}
\usepackage{graphicx}
\usepackage{textcomp}
\usepackage{xcolor}

\def\BibTeX{{\rm B\kern-.05em{\sc i\kern-.025em b}\kern-.08em
    T\kern-.1667em\lower.7ex\hbox{E}\kern-.125emX}}
\begin{document}

\title{Impact of Integrated Circuit Packaging on Synaptic Dynamics of Memristive Devices}

\author{\IEEEauthorblockN{Aidana Irmanova, Grant A.Ellis, Alex Pappachen James}
\IEEEauthorblockA{\textit{Department of Electrical and Computer Engineering} \\
\textit{Nazarbayev University}\\
Astana, Kazakhstan\\
Email: \{aidana.irmanova,grant.ellis,alex.james\}@nu.edu.kz}

}

\maketitle

\begin{abstract}

The memristor can be used as non volatile memory (NVM) and for emulating neuron behavior. It has the ability to switch between low resistance $R_{on}$ and high resistance values $R_{off}$, and exhibit the synaptic dynamic behaviour such as potentiation and depression. This paper presents a study on potentiation and depression of memristors in Quad Flat Pack. A comparison is drawn between the memristors with and without the impact of parasitics of packaging, using measured data and equivalent circuit models. The parameters in memristor and packaging models for the SPICE simulations were determined using  measured data to reflect the memristor parasitics in Quad Flat Packs. 


 

\end{abstract}


\section{Introduction}

Memristors are nonlinear devices that has the potential to drive the development of more than Moore's law integrated circuits. Single memristor itself can store multilevel values which is controlled by the voltage or current applied across the device as a function of time. Ideally, it can be programmed to any resistance level between a maximum possible resistance $R_{off}$ and minimum possible resistance $R_{on}$ \cite{kannan2015modeling}. Memristors have small area, fast read times, low leakage currents which makes it promising non-volatile memory device. They are also known to be useful in neuromorphic applications as they exhibit the dynamics similar to the time-dependent changes in synaptic connections that alter the synaptic weights in neural networks (NN). If the memristor is exposed to the same stimuli over and over again, the change in memristance value of devices decreases over time, which is similar to the learning process of neurons. This is called long-term potentiation or long-term depression \cite{ltp} of synapses.


Actual memristor devices are mathematically represented with extended memristor models. However, the existing models do not include switching stochasticity, which is important for extensive simulations of circuits prior to the fabrication. The realistic simulation of systems on chip with memristors require models that include memristance variations for $R_{on}$ and $R_{off}$ states and parasitics values associated with packaging that affects the circuit behaviour. As writing and erasing time of memristors are in nanoseconds range, the high frequency operation of the devices should be tested as well. 

The aim of this paper is to analyse the impact of memristance variability and packaging on memristive devices. The parameters of equivalent circuit  (R, L, C values)  of packaged memristor used in this work were selected to model bond wires used in Quad Flat packs or Ball grid arrays which are of the interest for the low cost solution. Despite the speed limitation of Quad Flat Packs, its design can still be modified to provide higher speed of operation as in\cite{sun2013design}


    
The remaining of the paper is organized as follows: Section II presents the details on the experimental setup of memristance measurements and memristor model modification. Section III introduces the equivalent circuit of packaged memristors and provides simulation results of both ideal outputs and the errors associated with a packaging effect.

\section{Determining Memristor Model Parameters}

There are different type of materials that memristors are fabricated such as Hafnium\cite{kumar2016conduction}, Titanium\cite{xia2009memristor}, and Tungsten \cite{ duraisamy2012fabrication}. Depending on the material stack and structure of memristor device the ratio of $R_{on}$ and $R_{off}$ values, switching times, and switching threshold values can vary dramatically. 
The read and write pulse voltages, and the ON time period needs to be carefully selected to observe the memristor characteristics. Figure 1 presents the experimental setup for measuring $R_{on}$ (a) and $R_{off}$ memristance values of the memristor\cite{campbell2015carbon} , that consists of (4) Keithley 4200 Parametric Analyzer\cite{testequity}, (3)   DIP chip, containing 8 discrete memristors, (2) twisted wires that connect (1) the test fixture 8101-4TRX with the source and measure units to the pins of the memristor in a package.

\begin{figure}[!ht]
\centering
   
    \includegraphics[height =6cm]{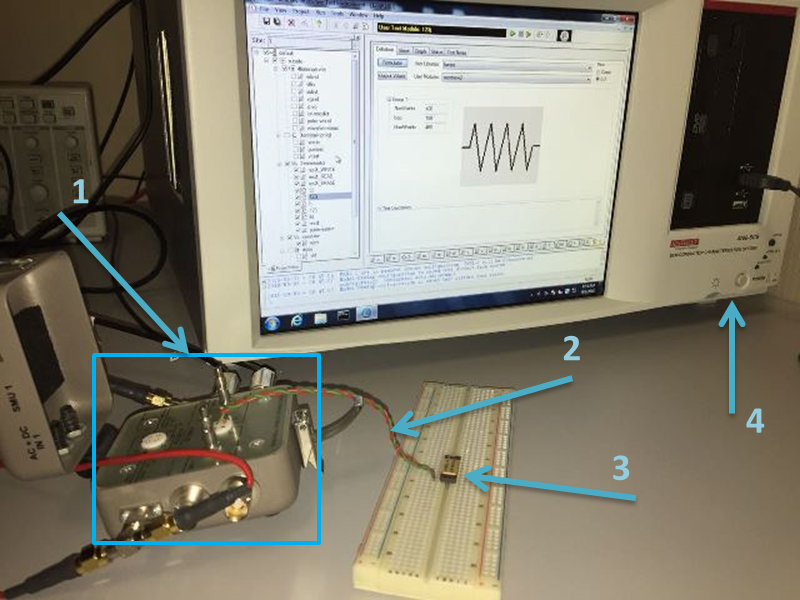}
    
     \caption{Experimental Setup: 1. Test fixture; 2. Twisted wires connecting the memristor pins to Source and Measure units; 3. DIP memristor chip; 4. Keithley 4200 Parametric analyzer.}
     \label{r}
     \end{figure}

\begin{figure}[!ht]
\centering
	\subfigure[]
    {
	\includegraphics[width = 4cm]{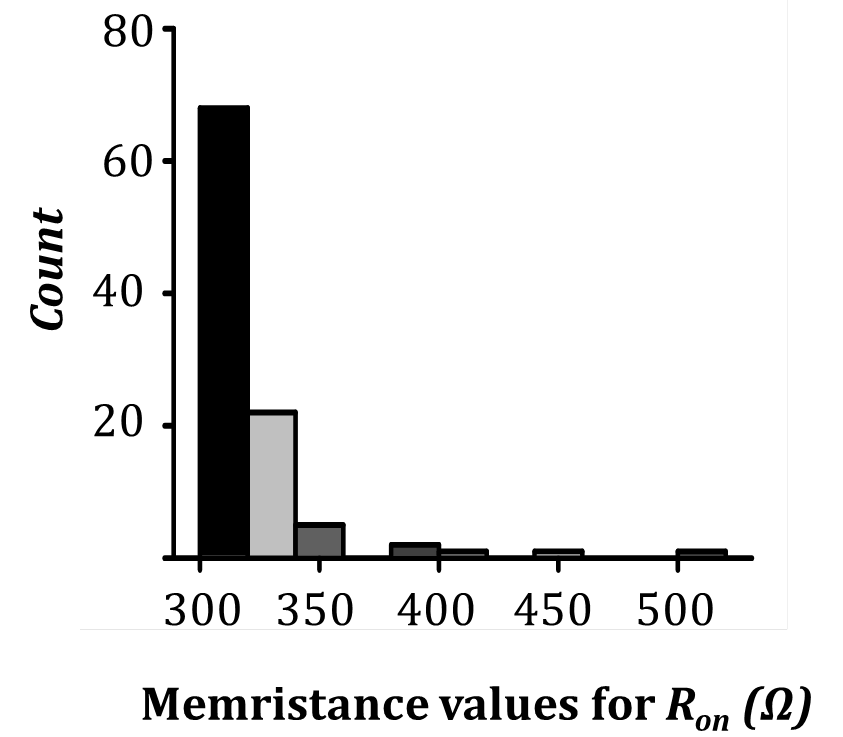} 
    }\subfigure[]
    {\includegraphics[width =4cm]{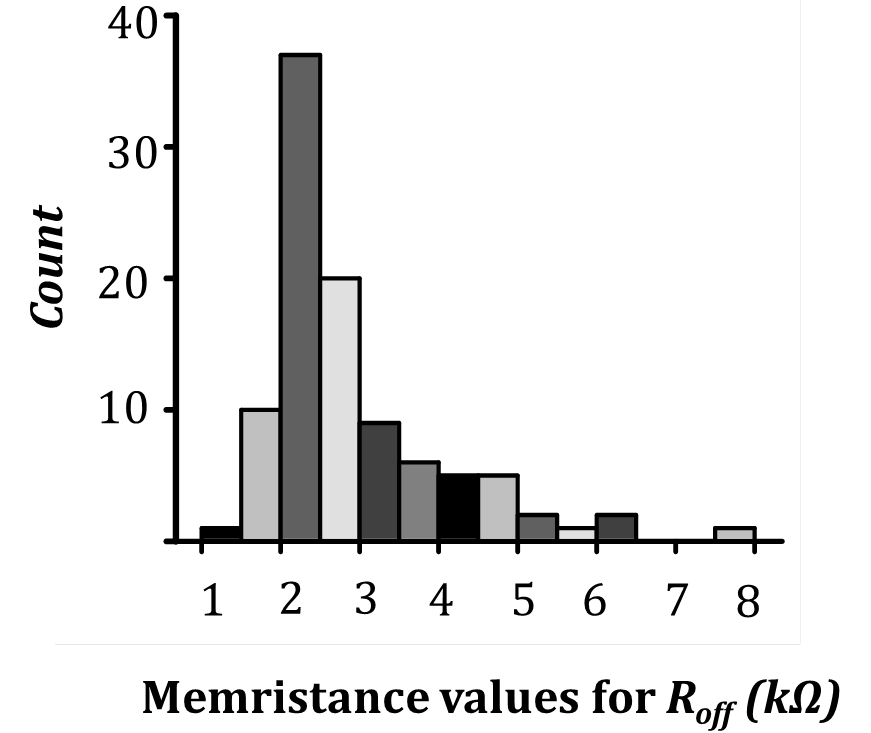} }
    
     \caption{Distribution of memristance values for $R_{on}$ (a) and $R_{off}$(b) }
     \label{r}
     \end{figure}

     \begin{figure}[!ht]
\centering
    \subfigure[]
    {
	\includegraphics[height =2
    cm]{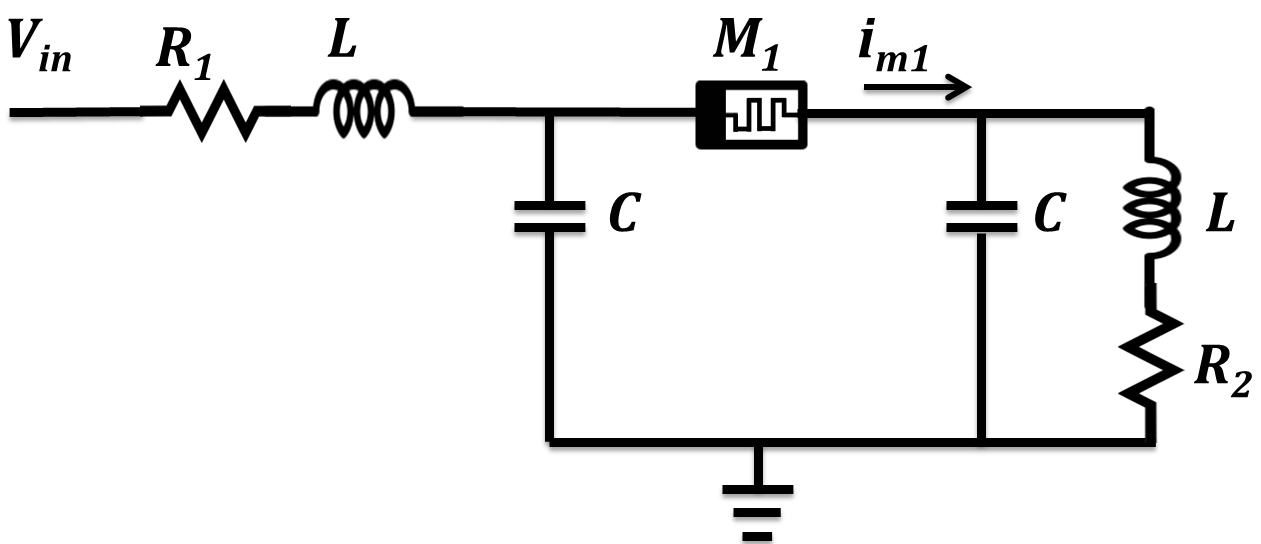} 
    }\subfigure[]
    {\includegraphics[height =2cm]{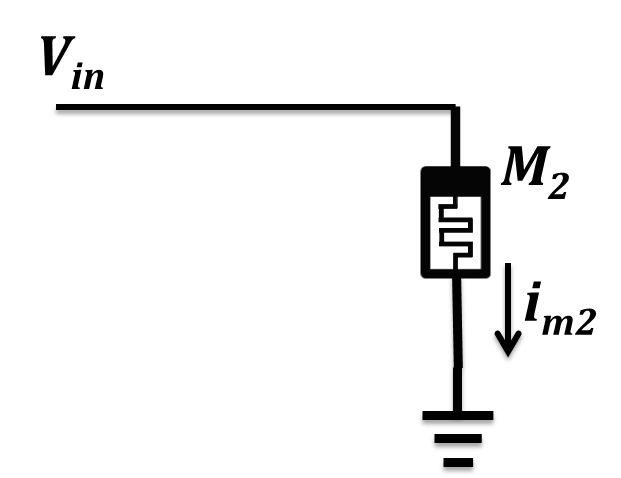} }
    
     \caption{Equivalent circuit model (a) and ideal circuit (b) for single cell of packaged memristor devices.}
     \label{r}
     \end{figure}

\begin{figure}[!ht]
\centering

	\subfigure[]
    {
	\includegraphics[height = 2.1cm]{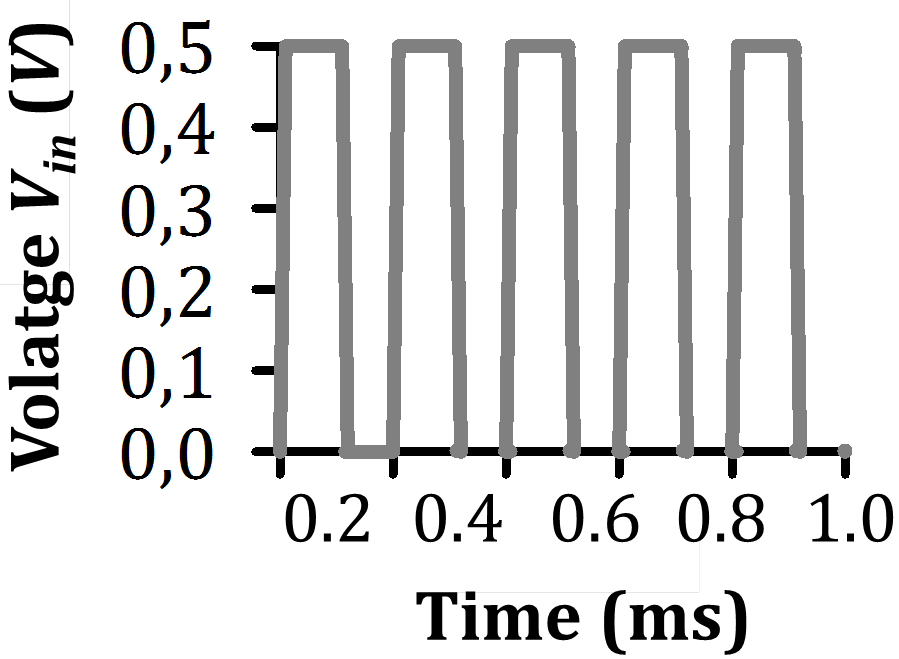} 
    }
    \subfigure[]{
	\includegraphics[height = 2.1cm]{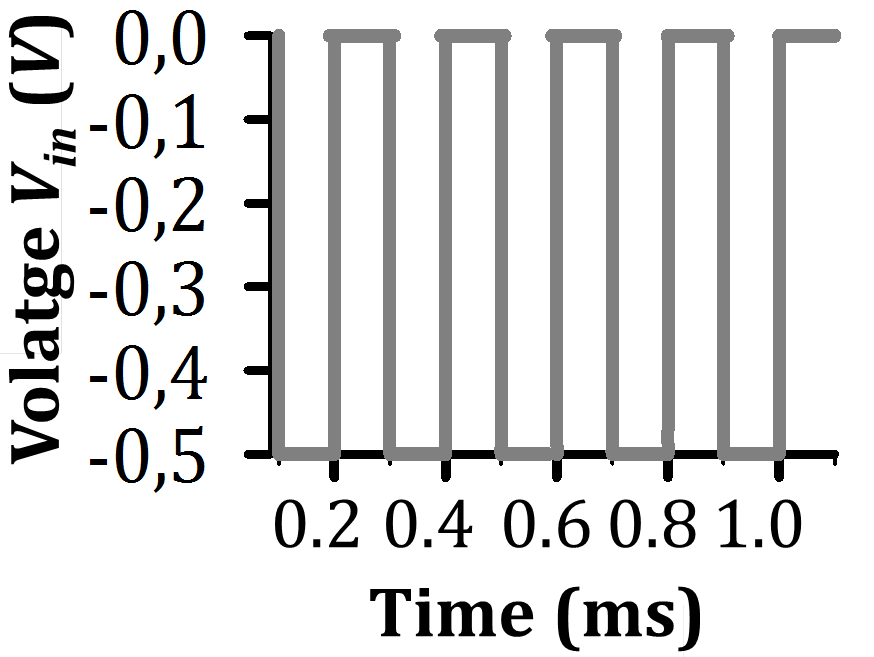} 
    }
    \subfigure[] {\includegraphics[height = 2.2cm]{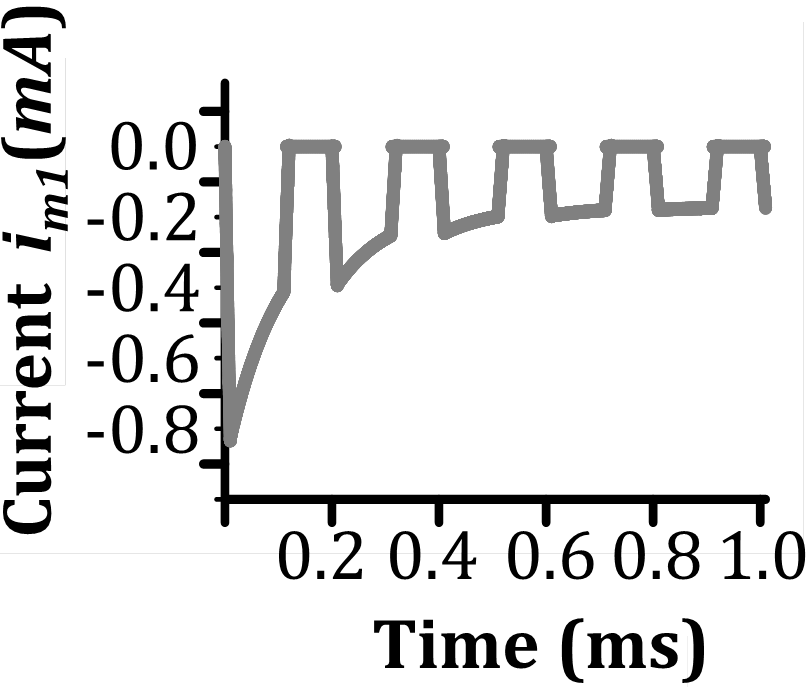}
    }
      \subfigure[] {\includegraphics[height = 2.2cm]{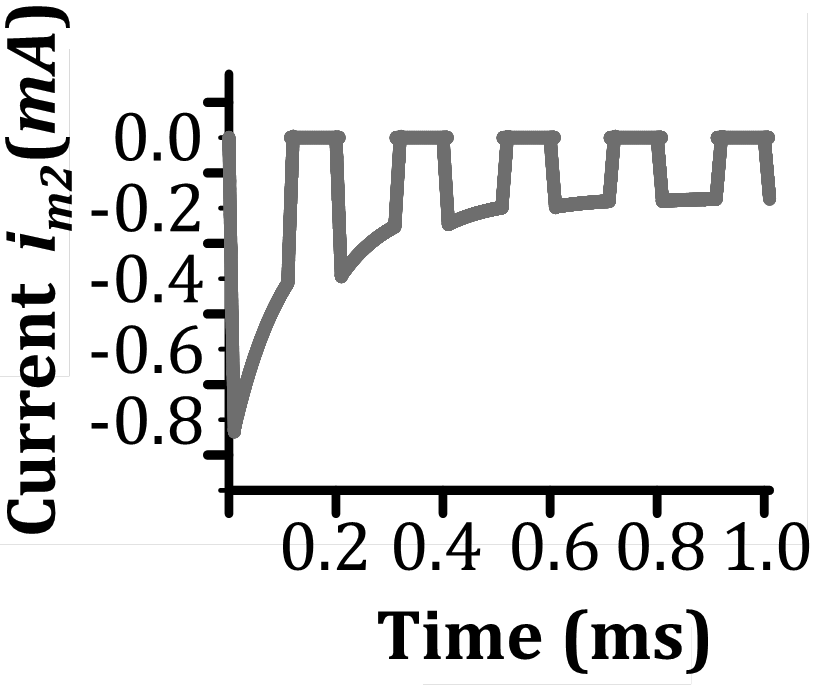}}   
    \subfigure[]{\includegraphics[height = 2.2cm]{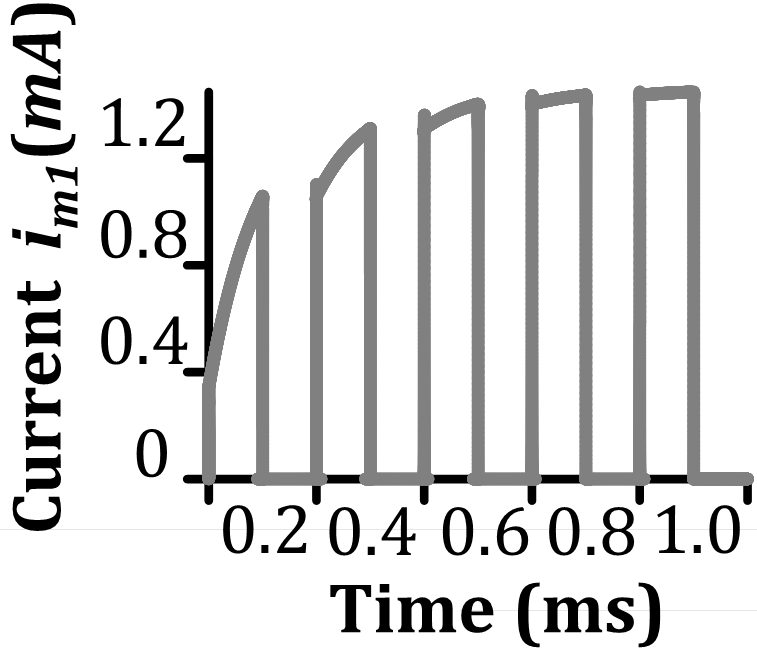}}
       \subfigure[]
    {\includegraphics[height = 2.2cm]{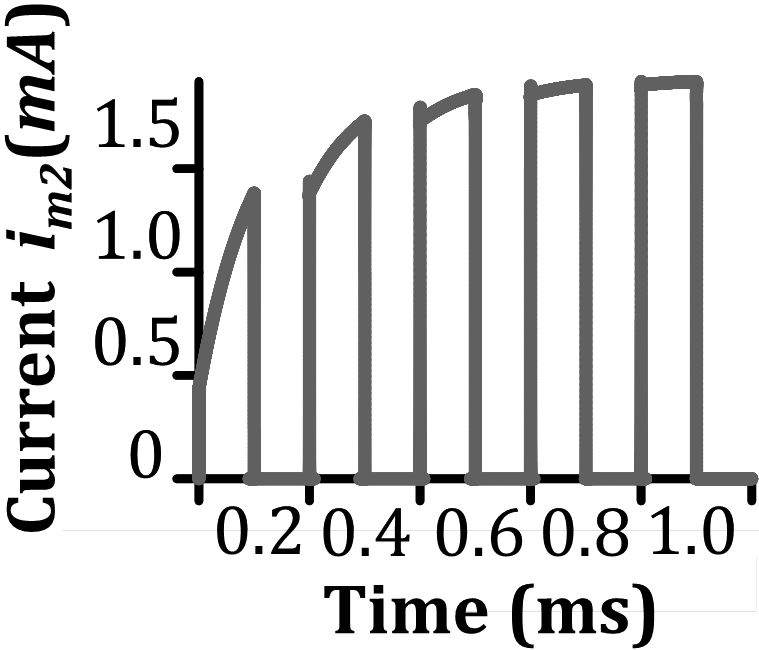}}
     \subfigure[]
     {\includegraphics[height = 2.2cm]{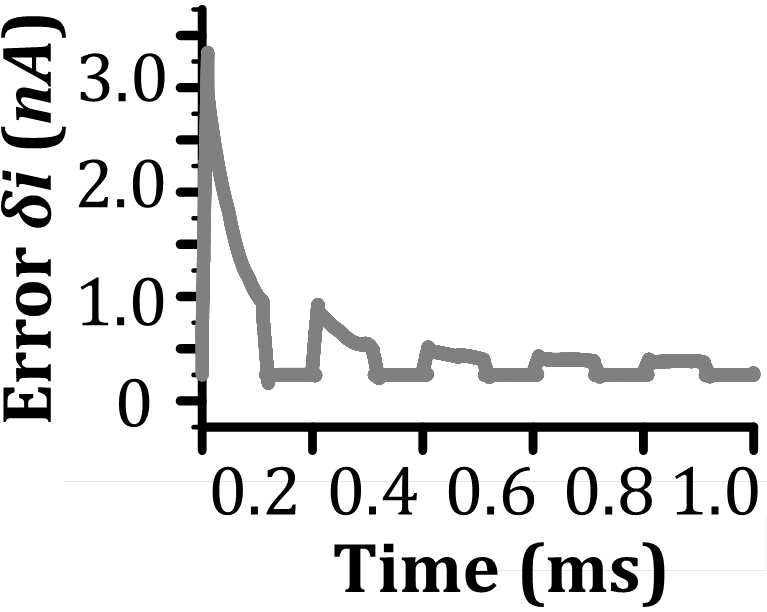} }
      \subfigure[]
     {\includegraphics[height = 2.2cm]{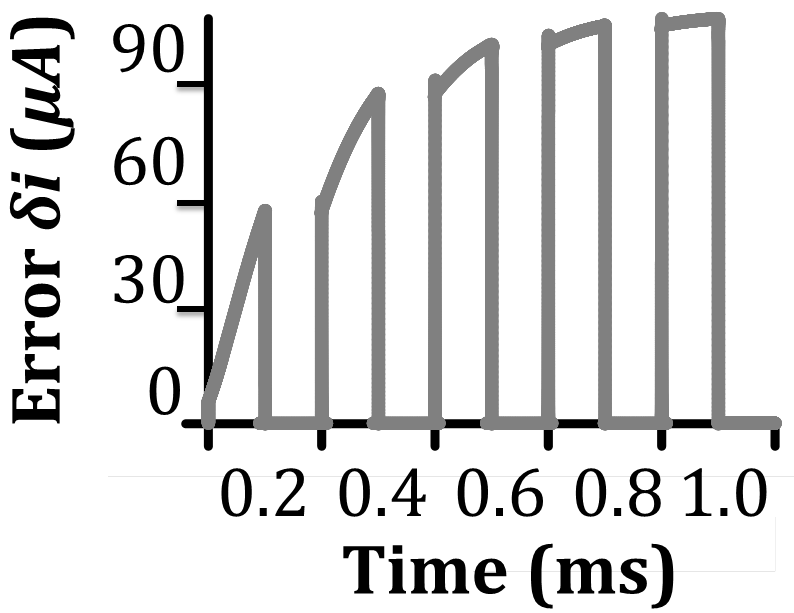} }

     \caption{Input and output values of the circuit using memristor model with $R_{on}$ =$R_{on}^{mean}$, $R_{off}$ = $R_{off}^{mean}$. (a) Voltage frequency is $10k V_{in}=0.5V$;  (b) Voltage frequency is $10k V_{in}=-0.5V$; (c, e) output current of packaged memristor circuit and (d,f) ideal circuit and (h,i) the $\delta i$ respective error for $ V_{in}=\pm 0.5V$. }
     \label{io}
     \end{figure}
          
\begin{figure}[!ht]
\centering
	\subfigure[]
    {
	\includegraphics[height = 3cm]{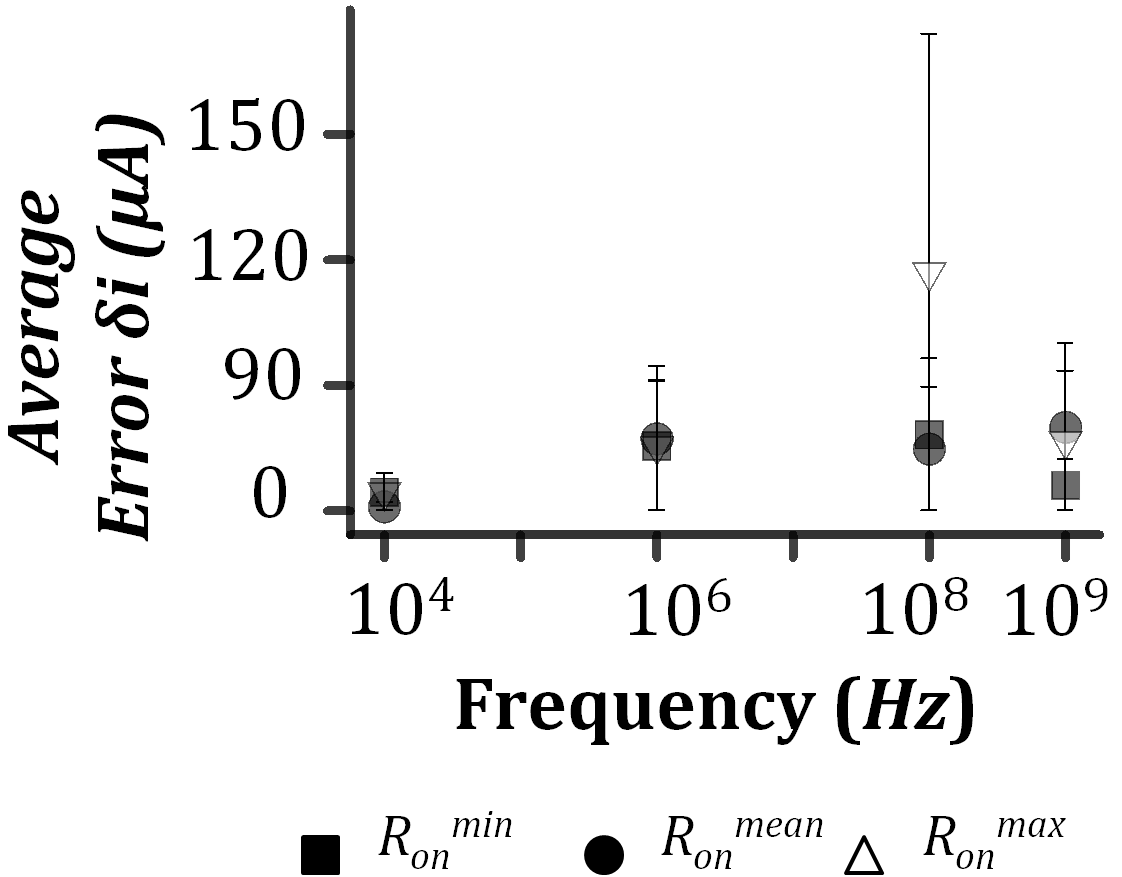} 
    }\subfigure[]
    {\includegraphics[height =3cm]{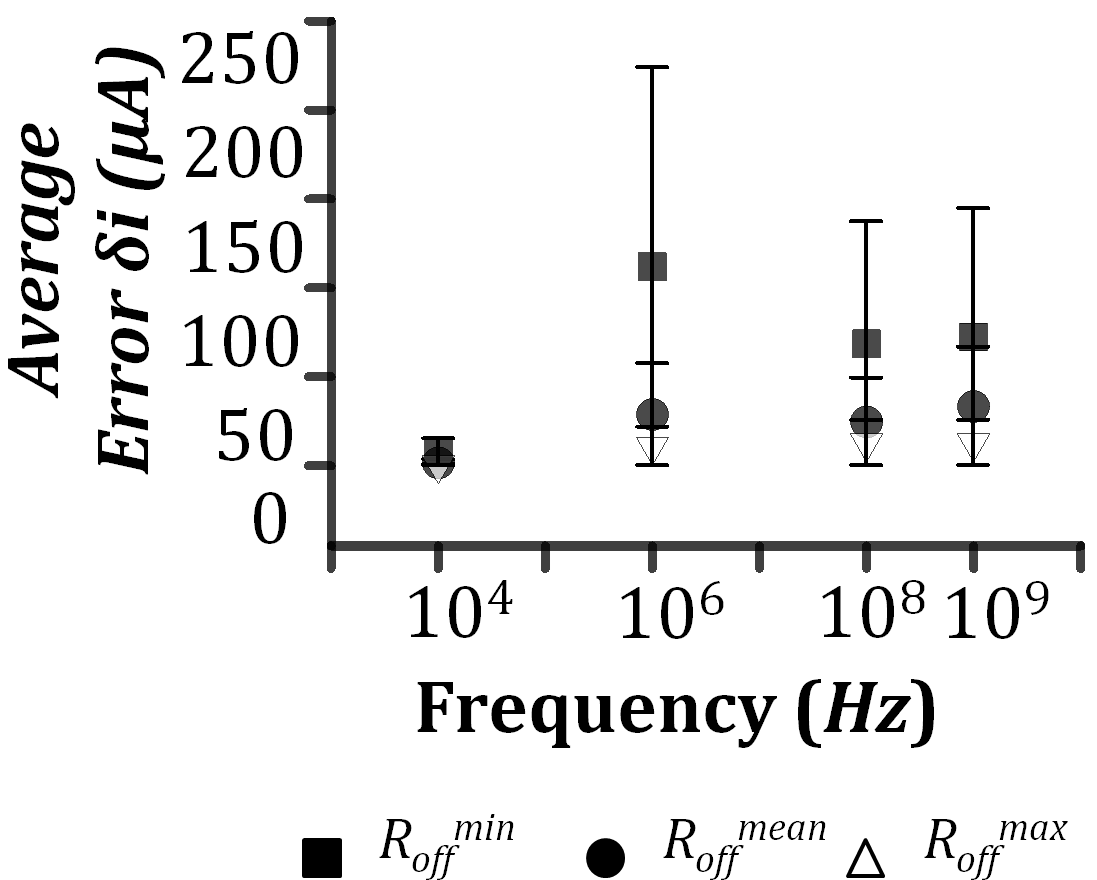} }
    
	\subfigure[]
    {
	\includegraphics[height = 3cm]{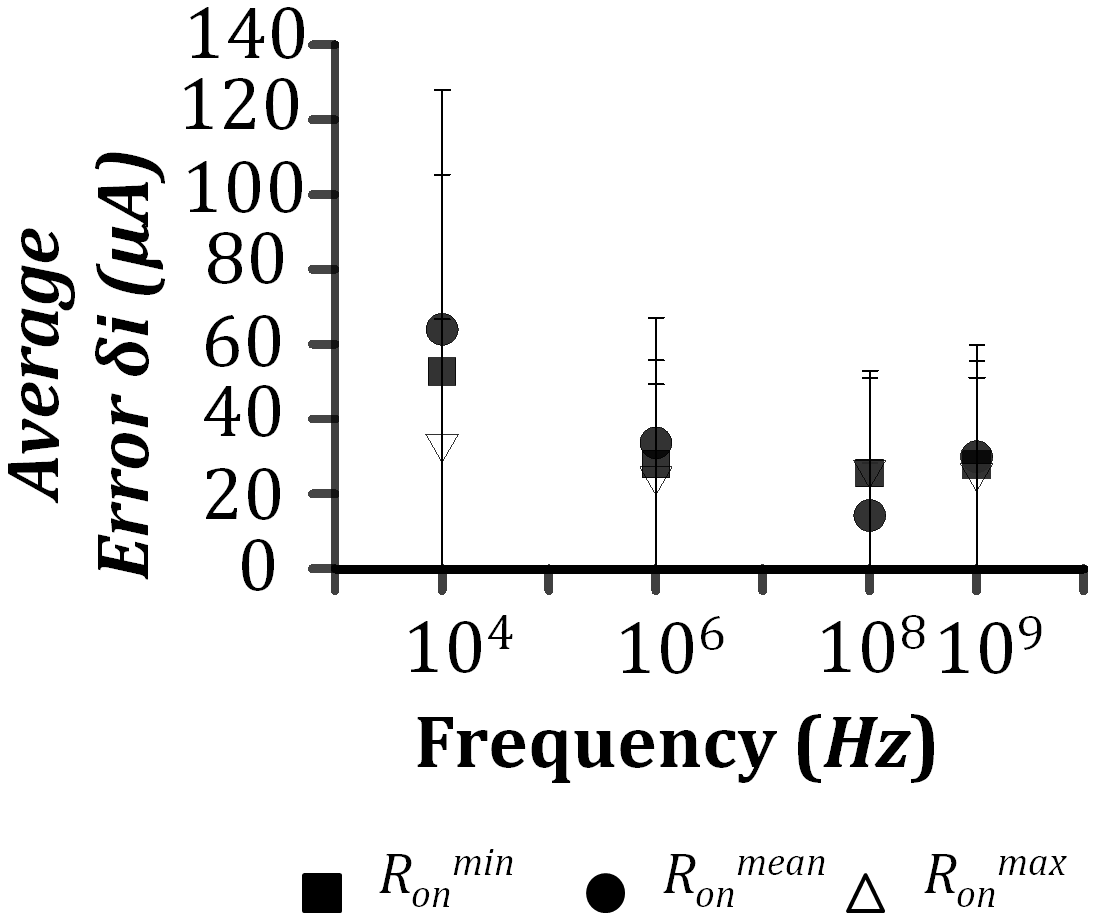} 
    }\subfigure[]
    {\includegraphics[height =3cm]{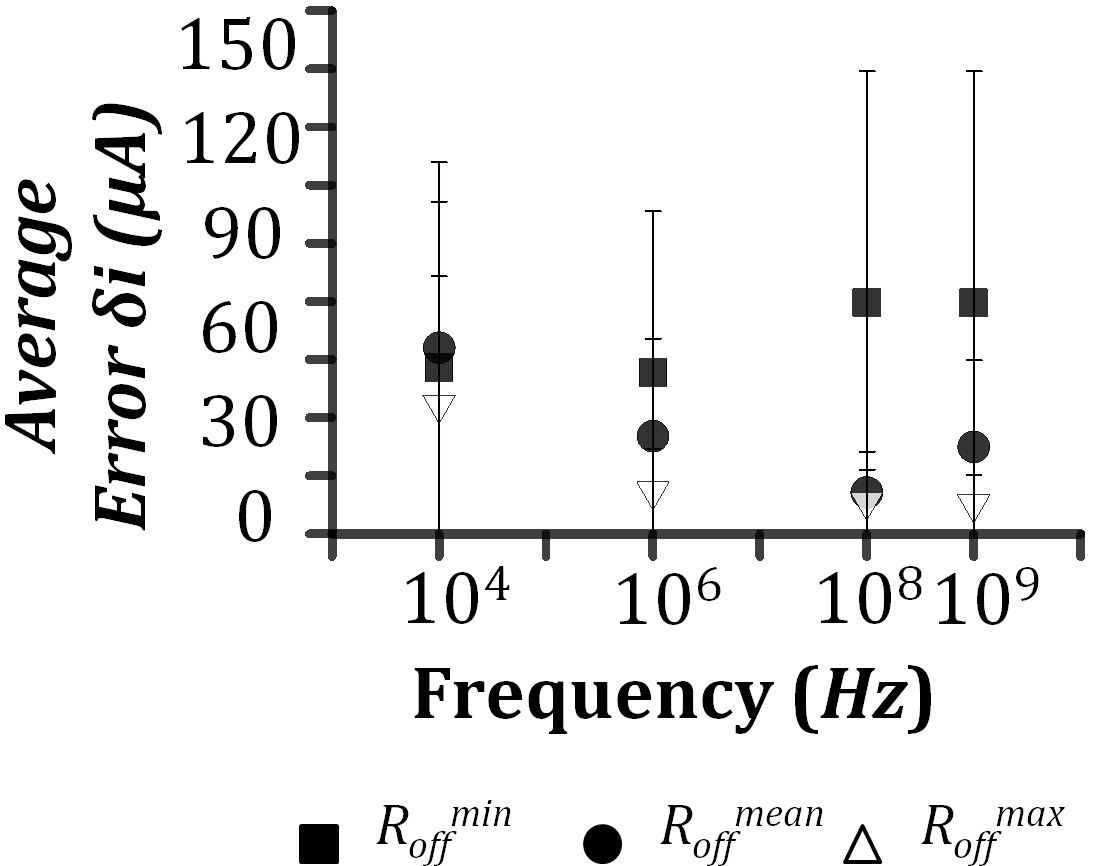} }

    \caption{Calculated average output current error with varying  (a) $R_{off}$ and  (b) $R_{on}$ values for  the range of operating frequencies (10KHZ, 1MHz, 100MHz, 1GHz) with $V_{in}= 0.5V$;Calculated average output current error with varying  (c) $R_{off}$ and  (d) $R_{on}$ values for  the range of operating frequencies (10KHZ, 1MHz, 100MHz, 1GHz) with $V_{in}=-0.5V$;   }
     \label{avge}
     \end{figure}

\subsection {Experimental setup and measurement results}

For this paper, memristance measurements of chalcogenide memristors \cite{campbell2015carbon} in 16-pin DIP package were performed for only two states that correspond to high resistance $R_{off}$ and low resistance $R_{on}$. The memristors used in the experimental measurements are sensitive to high voltage amplitudes and they best operate with stimuli of nanosecond width. The parametric analyzer shown in Fig.1  was used for generating square voltage pulses with 140ms width and  $\pm 0.5V$ amplitude. The data collected from the experiment was limited to the 100 writing and erasing cycles in order to avoid damaging the memristors.

 \begin{table}[!ht]
 \centering
 \caption{Measured Memristance values}
\begin{tabular}{|l|l|l|l|}
\hline
          & Mean ($\mu$)    & Max    & Min    \\ \hline
$R_{off}$ & 2924.655 & 7682.6 & 1432.4 \\ \hline
$R_{on}$  & 323.4    & 512.03 & 305.34 \\ \hline
\end{tabular}
\end{table} 
The memristor chips are in the early stages of its development and have large device variability in terms of $R_{on}$ and $R_{off}$ values and the switching thresholds.  For measuring the $R_{on}$ and $R_{off}$ variations of memristors, they have to be programmed to its respective state and read with the voltage lower than its switching threshold. In this paper, the memristance values were obtained altering the applied voltage to $+0.5V$ for writing and $-0.5V$ for erasing and $0.05V$ for reading. The switching threshold of memristor used in the experiment was $0.15V$. The histograms presented in Figure 2 shows distribution of memristance values for $R_{on}$ and $R_{off}$ states. Table 1 shows the maximum, minimum and mean values of $R_{on}$ and $R_{off}$ states.

\subsection{Modification of model parameters} 
For the simulation of packaged memristors the Knowm model \cite{knowm} was used. In this model the lowest $R_{on}$ and highest $R_{on}$ memristance states were initially defined as $500\Omega$ and $1500\Omega$ respectively.  The model parameters were modified replacing the $R_{on}$ and $R_{on}$ states with experimental data. The corner analysis and simulations are performed on five different combinations of memristor values, $\{R_{on}, R_{off}\}$, they are: (1)$\{R_{on}(\mu), R_{off}(\mu)\}$, (2) $\{R_{on}(\mu), R_{off}(\text{max})\}$, (3) $\{R_{on}(\mu), R_{off}(\text{min})\}$, (4) $\{R_{on}(\text{max}), R_{off}(\mu)\}$, and (5) $\{R_{on}(\text{min}), R_{off}(\mu)\}$.

\section{Analyzing the Error Associated with Package}
In this section, the equivalent model of packaged single memristor is presented and simulation results of both ideal and packaged memristor circuits are provided.

\subsection{Equivalent model of packaged memristor}

Using memristors in circuits involves designing additional programming circuitry for setting the memristance values to $R_{on}$, $R_{off}$ or other intermediate values. Depending on whether the memristor is at in programming mode (it is connected to the writing circuit at the input and shorted out at the output) or operating mode (it is connected to the readout circuit at the output) the equivalent circuit of interconnections between the memristive circuit and the electronic package will differ. 

Figure 3a shows the equivalent model of the packaged memristor, while Fig. 3b shows an ideal case.
The resistance values $R_1$ and $R_2$ represents the circuit blocks to which the memristor is connected.  At the input of memristor's positive terminal $R_1$ represents the resistance of the writing circuit with low impedance. The output of the memristor circuit can be shorted out with a low $R_2$ or connected to the high impedance readout circuit. For the generalization of the circuit, we used $R_1$ and $R_2$ as $10 \Omega$ and approximated Quad Flat Pack parameters for wire inductance $L = 1.2nH$ and capacitance to ground $C = 25fF$ (capacitance to the substrate is ignored) as in  \cite{unchwaniwala2001electrical}. 

\subsection{Simulation results and Error analysis}
In this section, the error associated with package parasitics is calculated for five cases of memristor resistance combinations as outlined in Section IIB. For the calculation of difference between ideal circuit output and packaged memristor circuit, a train of square voltage pulses $V_{in}$ with frequency in the range from $10k$Hz to $1G$Hz was used as an input to the circuit. Fig. 4 shows the example of $V_{in}$ at $10k$Hz (Fig.4a) and the corresponding current output of packaged memristor $i_{m1}$ (Fig.4c) and  memristor circuit without parasitic effect $i_{m2}$ (Fig.4d). Fig.4g shows the error $\delta i$ which is calculated as the difference between the current outputs  $i_{m2}$ and $i_{m1}$.

\begin{equation}
\delta i= i_{m1}-i_{m2}
\end{equation}

The graphs Fig. 4(c) and Fig. 4(d)  show the potentiation effect, while Fig. 4(e) and Fig. 4(f) show the depression effect of memristance values during the first millisecond (0-1ms). For further estimation of error variations, the average error was calculated for the first 5 cycles of $V_{in}$ with different frequencies. Figure 5 shows the overall trend of a substantial increase in the average error and its standard deviation with an increase in frequency and resistance values. It also shows the range of average error expected for variations of $R_{on}$ (Fig.5 a,c) and $R_{off}$ (Fig.5 b,d) values. For instance, driving the $0.5V$ to the  input of the circuit at frequency 10$k$Hz, the current output should be expected with the error ranging from $1.6\mu A$ to $7.5\mu A$ while at 1$G$Hz frequency the error range is larger, i.e., from $10 \mu A $ up to $72\mu A$. 

\section {Conclusion}
In this paper, we presented the impact of variability of memristances and Quad Flat Packs in the realistic memristive devices. The data collected from experimental measurements reveals that the synaptic dynamics of memristors are affected with packaging. Performed modifications and the equivalent model of parasitics provide extended evaluation of error associated with the packaging effect on memristor switching dynamics. Such analysis will be useful in development of realistic neural circuits and systems with memristors and memristor arrays.



\bibliographystyle{IEEEtran}
\bibliography{ref}

\end{document}